
%
\input phyzzx.tex
\overfullrule=0pt
%
%

\def\RREF#1#2{\gdef#1{\REF#1{#2}#1}}    
\def\jnl#1&#2(#3){\begingroup\let\jnl=\dummyj@urnal\sl #1\bf#2\rm
    (\afterassignment\j@ur\count255=#3)\endgroup}       
%
\def\PRL{ {\sl Phys. Rev. Lett.}   }

\def\NP { {\sl Nucl. Phys.}        }
\def\PL { {\sl Phys. Lett.}        }
\def\CMP { {\sl Comm. Math. Phys. }     }
\def\MPL { {\sl Mod. Phys. Lett.}        }
\RREF\brez{E. Brezin and V. Kazakov, \PL {\bf B236} (1990) 144.}
\RREF\shen{M. Douglas and S. Shenker, \NP {\bf B335} (1990) 135.}
\RREF\gros{D. Gross and A.A. Migdal, \PRL {\bf 64} (1990) 127;
\NP{\bf B340} (1990) 333. }
\RREF\doug{M. Douglas, \PL {\bf B238} (1990) 176. }
\RREF\davi{F. David, \NP {\bf B348} (1991) 507; \MPL
{\bf A5} (1990) 1019 }
\RREF\dall{S. Dalley, C. Johnson and T. Morris, \NP
{\bf B368} (1992) 655. }
\RREF\john{S. Dalley, C .Johnson and T. Morris,
\NP {\bf B368} (1992) 625. }
\RREF\peri{V. Periwal and D.  Shevitz, \PRL {\bf 64} (1990) 1326;
 \NP {\bf B344} (1990) 731  }
\RREF\seib{M. Douglas, N. Seiberg and S. Shenker \PL
{\bf B244} (1990) 381. }
\RREF\watt{S. Dalley, C. Johnson, T. Morris and A. Watterstam,
 Princeton, Southampton and Goteborg  preprint PUPT-1325,
SHEP 91/92-19 and ITP 92-20,
hepth@xxx/9206060. } \RREF\drin{V.G. Drinfeld and V.V. Sokolov, Sov. Jour.
Math. (1985)
1975.} \RREF\crnk{\u C. Crnkovi\'c, M. Douglas and G. Moore, Yale/Rutgers
preprint YCTP-P25-91 /RU-91-36  .}
\RREF\moor{G. Moore, \CMP {\bf 133 }(1990) 261; Prog. Theor. Phys. Suppl.
{\bf 102} (1990) 255. }
\RREF\mira{T. Hollowood, L. Miramontes, A. Pasquinucci and C. Nappi,
 \NP {\bf B373} (1992) 247 . }
\RREF\wils{B.A. Kupershmidt and G. Wilson ,  Invent. Math.
{\bf 62} (1981) 403. }
\RREF\spen{C. Johnson, T. Morris and B. Spence Southampton
(Imperial) preprint SHEP 90/91-30 (TP/91-92/01) . }
\RREF\kost{I.K. Kostov, \PL {\bf B238} (1990) 181. }
\RREF\crmo {\u C. Crnkovi\'c and G. Moore, \PL {\bf B257} (1991) 322.}
\RREF\gelf{I.M. Gelfand and L.A. Dikii, Russ. Math. Surv. {\bf 30} (1975)
77.}

\def\ULB{\address{Service de Physique Th\'eorique \break
Universit\'e Libre de Bruxelles, Boulevard du Triomphe \break
      CP 225, B-1050 Bruxelles, Belgium}}
\nopubblock
\titlepage
\line{\hfil\vbox{
\hbox{ULB--TH--06/92}
\hbox{August 1992}
}}

\title{2D Quantum Gravity and the Miura Map}

\author{L. Houart\footnote{\dag}{Aspirant FNRS.}\footnote{\sharp}{e-mail:
lhouart@ulb.ac.be}} \ULB
\abstract
{We study the $sL(3,C) \ mKDV$ string theories. We obtain the flows
and the string equations. Using the generalized Miura map, we show
that we have an ``unification'' of these models with the
$[\tilde P,Q]=Q \ sL(3,C) \ KDV$ ones in the framework of open-closed
string theories in minimal models backgrounds. }   \endpage

\chapter{INTRODUCTION}

Recently important progress has been made in the description of
theories which look very similar to two-dimensional gravity
coupled to matter. Starting with Hermitian matrix models in
the simplest phase, one finds that the $KDV$ integrable hierarchy
supplied by the condition $[P,Q]=1$ leads to a possible candidate
for 2D gravity coupled to minimal conformal models
\refmark{\brez,\shen,\gros, \doug}. However this approach doesn't
give a successful non-perturbative description for all the models;
in particular it's not the case for the simplest one : pure gravity
\refmark{\davi}.\hfill \break Another possible definition which is
based on the assumption that the $KDV$ flows hold non-perturbatively
has been proposed\refmark{\dall}. These theories are described by the
$KDV$ hierarchy supplied by the condition $[\tilde P,Q]=Q$ and can be
seen as arising from complex matrix models\refmark{\john}. This approach
leads to a successful non-perturbative description for all the models.

Apart from the theories based on the $KDV$ integrable hierarchy, there
exists
models giving rise in the double scaling limit to a description in terms of
the $mKDV$ hierarchy. They were first found by considering unitary matrix
models\refmark{\peri} then they were obtained in the double scaling limit
of
usual Hermitian matrix models in the 2 cuts phase\refmark{\seib}. Many
conjectures\refmark{\crnk} have been advanced in order to identify
physically these theories.\hfill \break Recently some authors
\refmark{\watt} argued that the $mKDV$ theories and the $KDV$ ones
supplied by the $[\tilde P,Q]=Q$ condition are two descriptions of
the same 2D gravitational system. They showed by studying the so-called
Miura map that the $sL(2,C)$ $mKDV$ string equation maps to the
$[\tilde P,Q]=Q$ one with a non-zero open string coupling constant.

This letter is concerned with a generalization of this identification. We
study the $mKDV$ theories associated with the $sL(3,C)$ algebra
(noted $mKDV(3)$) in the Zakharov-Shabat(ZS) formalism\refmark{\drin}.
We obtain the corresponding flows and string equations. After that using
the
generalized Miura map we show that the $mKDV(3)$ string equations are
mapped
onto the  $[\tilde P,Q]=Q$ string equations corresponding to the
Boussinesque
hierarchy with a non-zero open string coupling constant.

This paper is organized as follows : section 2 deals with a review of the
known results in the $sL(2,C)$ case. In section 3, we study the $sL(3,C)$
generalization. Finally, in section 4, we discuss the results.

\chapter{THE $sL(2,C)$ THEORIES}

In this section we review the connections existing between
the $mKDV$ string models and the $KDV$ ones via the Miura map.\hfill \break
In the ZS scheme the $mKDV$ flows are associated with the following first
order operator\refmark{\drin}: $$L_1 = \partial_x +{f \over 2}
{\pmatrix{1 & 0 \cr 0 & {-1} \cr}} + {\pmatrix{0 & \xi \cr 1 & 0 \cr}}
\eqn\eqi$$
Using the transformation $L_2 = S L_1 S^{-1}$ where :
$$ S = {\pmatrix { 1 & 0 \cr 0 &  \xi^{1 \over 2}}}\eqn\eqii$$
we can also describe the $mKDV$ flows with:
$$ L_2 = \partial_x + f {\sigma}_3 + \lambda {\sigma}_1 \eqn\eqiii$$
where we have $\lambda^2 = \xi$.\hfill\break
This is the operator $L$ we get when we study the 2 cuts Hermitian matrix
models characterized by even potentials\refmark{\moor,\crnk,\mira,\crmo}.
In
that context the physical specific heat is given by : $ (log Z)''=-{1 \over
4}
f^2 $ .\hfill \break
The flows compatible with the reduction \eqiii\ can be calculated by
recurrence and are given by \refmark{\crnk,\crmo}:
$$ {{df} \over {dt_k}}= F_{2k+1}\eqn\eqiv$$
where :
$$\eqalign{& H^{\prime}_k + f F_k = 0 \cr & G_{k+1} = F^{\prime}_k
+ f H_k \cr & F_{k+1} = G{\prime}_k \cr } \eqn\eqv$$
with the initial conditions : $F_0=0 \quad G_0=f \quad H_0=0$ and the
identification $t_0=x$.\hfill\break
The $mKDV$ massive string equation determined by the
compatibility\refmark{\moor} between the flows, $L_2$ and the operator
${d \over {d \lambda}}-M$ is :
$$\sum_{k=1} t_k(2k+1)G_{2k}+xf=0 \eqn\eqvi$$
The equations \eqiv ,\eqvi\ characterize completely the hierarchy of
the $mKDV$ models.

With the conventions choosen here the Miura map, which transforms the
$mKDV$ hierarchy into the $KDV$ one, is :
$$u={{f^2} \over 4} + {{f \prime} \over 2} \eqn\eqvii$$
Under \eqvii\ the flows \eqiv\ become :
$${{du} \over {dt_k}}= \partial_x [{1 \over 2} (F_{2k+1}-H_{2k+1})]
\eqn\eqviii$$
Defining
$$R_{k+1}[u]={1 \over 2}  (F_{2k+1}-H_{2k+1}) \eqn\eqix$$
Using \eqv , it's easy to check that :
$$R_{k+1}^\prime = R_k^{\prime \prime \prime} -4uR_k^\prime
-2 u^\prime R_k \eqn\eqx$$
Thus the $mKDV$ flows are indeed mapped onto the $KDV$ ones :
$${{du} \over {dt_k}} = \partial_x R_{k+1}[u] \eqn\eqxi$$
where the $R_k$ are the usual Gelfand Dikii potential\refmark{\gelf}
(with $ R_0=-{1 \over 2}$).

We now turn to the string equation. Using \eqix\ and \eqv\ we have :
$$G_{2k}=2D^\ast R_k \eqn\eqxii$$
where : $D^\ast= \partial_x -f$ \hfill \break
We can thus rewrite \eqvi\ as :
$$2D^\ast({\cal R})+1=0 \eqn\eqxiii$$
where :
$$ {\cal R}=\sum_{k=0} t_k(2k+1)R_k[u] \eqn\eqxiv$$
For a fixed critical model characterized by k we have : ${\cal R}=
R_k-{x \over 2}$.\hfill \break
It's now possible to extract $f$ :
$$f={{\cal R}^\prime + {1 \over 2} \over {\cal R}} \eqn\eqxv$$
Using the Miura map\eqvii\ we finally find :
$$({\cal R}^\prime)^2+4u {\cal R}^2-2{\cal R} {\cal R}^{\prime \prime}
={1 \over 4} \eqn\eqxvi$$
which is the $[\tilde P,Q]=Q \  KDV$ string equation\refmark{\dall} with a
non-vanishing open string coupling constant\refmark{\kost}.\hfill \break
This result argues thus for an unification of the $KDV$ and the
$mKDV \ sL(2,C)$ theories in the framework of open-closed string theory
in the $(2,2k-1)$ minimal models backgrounds\refmark{\watt}.

\chapter{The $sL(3,C)$ Generalized $mKDV$ MODELS}

We consider the $sL(3,C)$ algebra given by the following $3 \times 3$
matrices :\hfill \break
\settabs 4 \columns \+$S_1= \delta_{1,2}$ & $S_2= \delta_{1,3}$ & $S_3=
\delta_{2,1}$ & $S_4= \delta_{2,3}$ \cr
\+$S_5= \delta_{3,1}$ & $S_6= \delta_{3,2}$ & $H_1=diag(1,-1,0)$ &
$H_2=diag(1,0,-1) $ \cr

The generalization of \eqi\ is given by :
$$L_1 = \partial_x + {\pmatrix{q_1 & 0 & 0 \cr 0 & q_2 & 0 \cr 0 & 0
& q_3 \cr}} + {\pmatrix{0 & 0 & \xi \cr 1 & 0 & 0 \cr 0 & 1 & 0 \cr}}
\eqn\eqqi$$
where :
$$\sum q_i=0\eqn\eqqii$$
Performing the transformation $L_2=SL_1S^{-1}$ where $S= diag(1,
\xi^{1 \over 3},\xi^{2 \over 3})$ and using the condition \eqqii\
we write :
$$L_2= \partial_x + f H_1 +g H_2 + \lambda A_1 \eqn\eqqiii$$
where :
$$A_1=S_2+S_3+S_6\eqn\eqqiv$$
and ${\lambda}^3= \xi$

In analogy with \eqiii\ we are going to study the "string theory"
associated with the reduction \eqqiii .\hfill \break
We have first to find the flows coherent with this reduction. By the
ZS method for lie algebras\refmark{\drin}, we know that all the possible
candidates are : $${d \over d \tau_{i,k}} L_2=-[(M_{i,k})_+,L_2]=
[res(M_{i,k}),A_1]\eqn\eqqv$$
with $k=1,2,...$ and $i=1,2$;\hfill \break
where we have $M_{i,k}=e^{-adU}(A_i \lambda^k)$, the subscript $+$
stands for the part of the series with positive powers of
$\lambda $, and the transformation $e^{adU}$ is defined as usually
\refmark{\drin} by the fact that: $$\eqalign{e^{adU}(L_2) &=
L_2+[u,L_2]+{1 \over 2}[u,[u,L_2]]+...\cr & = \partial_x +\lambda
A_1+\sum^\infty_{i=0}(B_iA_1+C_iA_2) \lambda^{-i} \cr } \eqn\eqqv$$
where $u$ is a series of negative powers of $\lambda$ with coefficients
being functions of $x$ with values in $sL(3,C)$, and$A_2$ is the second
element of $Ker A_1$ : $$A_2=S_1+S_4+S_5 \eqn\eqqvi$$
Writing :
$$ Res(M_{i,k})=\sum^6_{j=1} a_{i,j,k}S_j + b_{i,k} H_1
+ c_{i,k} H_2 \eqn\eqqvii$$

we have the following recursion relations :
$$\eqalign{a_{i,3,k+1}-a_{i,2,k+1} &= a_{i,4,k}^\prime + (g-f) a_{i,4,k}
\cr
a_{i,6,k+1}-a_{i,3,k+1} & = a_{i,5,k}^\prime - (f+2g) a_{i,5,k}  \cr
b_{i,k+1}
+ 2 b_{i,k+1} &= a_{i,2,k}^\prime + (f+2g) a_{i,2,k} \cr b_{i,k+1}
- b_{i,k+1} &= a_{i,6,k}^\prime + (f-g) a_{i,6,k} \cr
a_{i,1,k+1}-a_{i,4,k+1}
&= b_{i,k}^\prime \cr a_{i,4,k+1}-a_{i,5,k+1} &= c_{i,k}^\prime \cr
a_{i,1,k}^\prime + a_{i,4,k}^\prime +  a_{i,5,k}^\prime &+ (2f+g)
a_{i,1,k}
 + (g-f) a_{i,4,k}-(f+2g)  a_{i,5,k} = 0 \cr a_{i,2,k}^\prime +
a_{i,3,k}^\prime + a_{i,6,k}^\prime &+ (f+2g)  a_{i,2,k} - (2f+g)
a_{i,3,k}+ (f-g) a_{i,6,k} = 0 \cr } \eqn\eqqviii$$
with the following non-zero initial values :
$$b_{1,0}=f \qquad c_{1,0}=g \qquad a_{2,2,0}=f \qquad a_{2,3,0}=g \qquad
a_{2,6,0}=-f-g \eqn\eqqix$$
Since the residues coherent with the reduction we are considering must
satisfy :
 $$b_{i,k}=c_{i,k}=a_{i,2,k}=a_{i,3,k}=a_{i,6,k}=0$$
 we have to restrict ourself to the following flows: $t_{i,k}
\equiv \tau_{i,i+3k}$ where $ k=0,1,2,... $ and \break
$i=1,2 $. These ones are thus given by :
$$\eqalign{{df \over dt_{i,k}}  &= \partial_x b_{i,i+3k-1}  \cr
{dg \over dt_{i,k}}  &= \partial_x c_{i,i+3k-1} \cr}\eqn\eqqx$$
where we have $t_{1,0}=x$.

We now turn to the determination of the string equations by flatness
conditions
\refmark{\moor}.\hfill \break
 The compatibility condition between $L_2$ and the operator
${d\over d \lambda} - P$ gives:
$[P,L_2]=A_1$.\hfill \break The solutions are given by
$P_{i,k}=(M_{i,i+3k-1})_+-xA_1$
which lead to :
$$\eqalign{b_{i,i+3k-1} &= xf \cr c_{i,i+3k-1} &= xg \cr} \eqn\eqqxi$$
The $mKDV(3)$ ``string theories'' are thus characterized by the flows
\eqqx\ and the string equations \eqqxi .

We now study the generalized Miura transformation. This one is defined
by the action of an upper triangular matrix $S_m$ on the $L_1$ operator
so that we have :
$$L_{kdv}=S^{-1}_mL_1S_m = \partial_x + {\pmatrix{ 0 & 0 & -({3 \over 4}
u_2^\prime +u_3) \cr 0 & 0 & -{3 \over 2} u_2 \cr 0 & 0 & 0 \cr}} +
{\pmatrix{0 & 0 & \xi \cr 1 & 0 & 0 \cr 0 & 1 & 0 \cr}} \eqn\eqqxii$$
which is the ZS formulation of the usual $sL(3,C)$ operator (A.1).
\hfill \break
The unique $S_m$ leading to \eqqxii\ gives us the following relations :
$$\eqalign{u_2 &= - {2 \over 3}(f^\prime +2g^\prime +f^2+g^2+fg)  \cr
u_3 &= -({f^{\prime \prime} \over 2} +ff^\prime +fg(f+g)+ {3 \over 2}
f^\prime g +{fg^\prime \over 2}) \cr} \eqn\eqqxiii$$
under the Miura map the flows become :
$${d \over {dt_{k,i}}} {\pmatrix{{3 \over 2} u_2 \cr u_3 \cr}}=
{\cal B}{\pmatrix{ f_k \cr g_k \cr}} \equiv
{\pmatrix{ 0 & \partial_x \cr \partial_x & 0 \cr}}
{\pmatrix{ R^2_{i,k+1} \cr R^3_{i,k+1} \cr}}\eqn\eqqxiv$$
 where we have defined $R^2_{i,k}$, $ R^3_{i,k}$  and
${\cal B}$ is the ``Frechet Jacobian'': $${\cal B}=
-{\pmatrix{ \partial +2f+g & 2 \partial +f+2g  \cr {\partial^2 \over 2}
+f \partial +{3 \over 2}g \partial +2fg+g^2+f^\prime +
{g^\prime \over 2} & {f \partial \over2}+{3 \over 2}
f^\prime +2fg+f^2 \cr}}\eqn\eqqxv$$
Using \eqqx , \eqqxiv\ and \eqqviii\ we get :
$$\eqalign{R^2_{i,k+1} &= {3 \over 2}f a_{i,5,i+3k}+
{3 \over 2}(a_{i,6,i+3k+1}+ a_{i,3,i+3k+1}) \cr
R^3_{i,k+1} &= 3a_{i,5,i+3k} \cr} \eqn\eqqxvi$$
By \eqqix\ we have the initial values :
\hfill \break  $R^3_{1,1}={3 \over 2}u_2 \quad R^3_{2,1}=2u_3
\quad R^2_{1,1}=u_3 \quad R^2_{2,1}=-{1 \over 4}
(u^{\prime \prime}_2+3u^2_2)$.\hfill \break
Using \eqqviii\  it's easy to show that the functions \eqqxvi\
satisfy the recursion relations (A.2) of the Boussinesque hierarchy
given in the appendix.
The transformation \eqqxiii\ maps thus, as expected\refmark{\wils},
the $mKDV(3)$ flows onto the $KDV(3)$ ones.

We now consider the string equation \eqqxi .
Taking its derivative, using \eqqx\ and \eqqxiv\ we get :
$${\cal B}{\pmatrix{ (xf)^\prime \cr (xg)^\prime }}=
{\cal D}_2{\pmatrix{  R^2_{i,k} \cr R^3_{i,k} \cr}}\eqn\eqqxvii$$
where we have used ${\cal D}_2$, the second hamiltonian structure of the
Boussinesque hierarchy (see appendix) .\hfill \break
Remarking that :
$${\cal B}{\pmatrix{ (xf)^\prime \cr (xg)^\prime }}=
{\pmatrix{  3u_2+{3 \over 2}xu^\prime_2 \cr 3u_3+xu^\prime_3 \cr}}
\eqn\eqqxviii$$
The string equation becomes :
$${\cal D}_2{\pmatrix{ {\cal R}_{2,i,k} \cr
{\cal R}_{3,i,k} \cr}}=0\eqn\eqqxiv$$
where we have : ${\cal R}_{2,i,k}= R^2_{i,k}-3x\ \hbox{and}
{\cal R}_{3,i,k}= R^3_{i,k}$ \hfill
\break
Now by multiplying on the left \eqqxiv\ by $({\cal R}_2,{\cal R}_3)$
and integrating once we finally get :
$$\eqalign{ & {1 \over 3}({\cal R}^{\prime}_2)^2-{1 \over 2}
u_2{\cal R}^2_2-{2 \over 3}{\cal R}_2{\cal R}^{\prime \prime}_2-
u_3{\cal R}_2{\cal R}_3 +{1 \over 18} \lbrace {\cal R}_3{\cal R}^{(4)}_3
-{\cal R}^\prime_3{\cal R}^{(3)}_3-{1\over 2}({\cal R}^{\prime \prime
}_3)^2
 \rbrace  \cr &+{5 \over 12} \lbrace u_2{\cal R}_3{\cal R}^{\prime
\prime}_3
 -{1 \over 2}u_2({\cal R}^{\prime}_3)^2+{1 \over 2}u^\prime_2
{\cal R}_3{\cal R}^\prime_3 \rbrace + {1 \over 12} \lbrace 3u^2_2+
u^{\prime \prime }_2 \rbrace {\cal R}^2_3=3 \cr} \eqn\eqqxv$$
The constant on the rhs of \eqqxv\ is determined by the scaling
property of the string equation \eqqxi\ . Indeed, defining
$f= \alpha {\tilde f} ,\  g= \alpha {\tilde g} \ \hbox{and}
\  {\tilde x}=\alpha x$ we have in the $\alpha =0$ limit
${\tilde f}={\tilde g}=0$. Thus we have ${\tilde u}_2={\alpha }^{-2}
u_2=0$ and ${\tilde u}_3={\alpha}^{-3} u_3=0$ which fix the constant to be
3.

We have thus shown that the Miura map transforms the $mKDV(3)$ string
equations \eqqxi\ onto \eqqxv\ which are the $[{\tilde P},Q]=Q$ equations
for the $(p,3) \ KDV$ models\refmark{\spen} with a non-vanishing constant
which can play the role of an open string  coupling\refmark{\kost,\watt}.

\endpage

 \chapter{DISCUSSION}

We have studied the $sL(3,C)$ generalization of the $mKDV$ string models
deriving the flows and the string equations. We have shown, generalizing
the results of Dalley et al.\refmark{\watt},  that the $KDV(3)$ and the
$mKDV(3)$ theories are unified by the Miura map in the picture of
open-closed
string theories in $(p,3)$ minimal backgrounds : the $mKDV(3)$ closed
string
equations being mapped onto the $KDV(3)$ open ones. It's natural to
conjecture
that we should obtain the same results for all $sL(n,C)$ models.

Finally there is an important question we would like to emphasize.
For the $sL(2,C)$ \break
$mKDV$ theory {--} the only one up to now which possesses an underlying
matrix model leading to an identification of the physical specific
heat {--} we see that the quantity which is solution of the $KDV
\quad [{\tilde P},Q]=Q$ equation after having performed the Miura map
is not the specific heat$-{f^2 \over 4}$ but $-{f^2 \over 4}-
{f^\prime \over 2}$. It would be interesting to understand this
discrepancy which could be related to the fact that we are mapping
a closed string theory onto an open one. We hope to address this
problem elsewhere.

\vskip 5 cm

\ack

I would like to thank \u C. Crnkovi\'c for valuable discussions.

\endpage

\appendix

The Boussinesque hierarchy is the one associated with :
$$Q=d^3+{3 \over 4} \lbrace u_2,d \rbrace +u_3\eqn\eqai$$
where $d \equiv \partial_x$

The corresponding flows are :
$$\alpha_i {\partial \over \partial t_{l,k}}u_i=
{\cal D}^{ij}_1 R^j_{l,k+1}={\cal D}^{ij}_2 R^j_{l,k}\eqn\eqaii$$
where $i,j=2,3\qquad \alpha_2={3 \over 2} \quad \alpha_3=1$ and
${\cal D}_1 \quad (\hbox{resp.} {\cal D}_2)$ is associated with
the first (resp.second) hamiltonian structure :
$${\cal D}^{22}_1={\cal D}^{33}_1=0 \qquad {\cal D}^{32}_1=
{\cal D}^{23}_1=d$$
 $${\cal D}^{22}_2={2 \over 3}d^3+u_2d+{1 \over 2}u^\prime_2 \qquad
{\cal D}^{23}_2=u_3d+{2  \over 3}u^\prime_3 \qquad {\cal D}^{32}_2=
u_3d+{1 \over 3}u^\prime_3$$
$${\cal D}^{33}_2=-{1 \over 18}d^5-{5 \over 12}u_2d^3-{5 \over 8}
u^\prime_2d^2-({1 \over 2}u^2_2+{3 \over 8}u^{\prime \prime}_2)d-
({1 \over 2}u^\prime_2 u_2 +{1 \over 12}u^{\prime \prime \prime}_2) $$
 And the first $R$ are : $$ R^2_{1,0}=3 \qquad  R^2_{2,0}=0
\qquad R^3_{1,0}=0 \qquad R^3_{2,0}=3$$

\endpage

\refout
\endpage
\bye